\documentclass[structabstract]{aa}  
\usepackage{fancyvrb}
\usepackage{multirow}
\usepackage{graphicx}
\usepackage{txfonts}
\begin{document}
\VerbatimFootnotes
   \title{Constraining the ages of the fireballs in the wake of the dIrr galaxy VCC1217 / IC3418}

   \author{Mattia Fumagalli \inst{1,2}, Giuseppe Gavazzi \inst{2},
          Roberto Scaramella \inst{3}, Paolo Franzetti \inst{4}
          }
   \institute{
   Leiden Observatory, Leiden University, P.O. Box 9513, 2300 RA Leiden, The Netherlands.\\
             \email{fumagalli@strw.leidenuniv.nl} 
	     \and
   	    Dipartimento di Fisica G. Occhialini, Universit\`a di Milano- Bicocca, Piazza della Scienza 3, 20126 Milano, Italy\\
             \email{giuseppe.gavazzi@mib.infn.it} 
	\and  INAF, Osservatorio Astronomico di Roma, via Frascati 33, 0040 Monte Porzio Catone (RM), Italy \\ 
             \email{roberto.scaramella@oa-roma.inaf.it}
	\and  INAF, IASF-Milano, Via Bassini 15, I-20133, Milano, Italy \\ 
             \email{paolo@lambrate.inaf.it}
   	     }
              \date{Received ...; accepted ...}

 
  \abstract
   {A complex of H$\alpha$ emitting blobs with strong FUV excess is associated to the dIrr galaxy VCC1217 / IC3418 
   (Hester et al. 2010), and extends up to 17 Kpc in the South-East direction.
   These outstanding features can be morphologically divided into diffuse filaments and compact knots, where most
   of the star formation activity traced by H$\alpha$ takes place.}
   {
   We investigate the properties of the galaxy and the blobs using a multiwavelength approach in order to constrain their origin.
   }
   {
   We collect publicly available data in UV and H$\alpha$ and observe the scene in the optical U,g,r,i bands with LBT.
   The photometric data allows to evaluate the star formation rate and to perform a SED fitting separately of 
   the galaxy and the blobs in order to constrain their stellar
   population age. Moreover we analyze the color and luminosity profile of the galaxy and its spectrum to 
   investigate its recent interaction with the Virgo cluster.
   }
   {
   Our analysis confirms that the most plausible mechanism for the formation of the blobs is ram pressure 
   stripping by the Virgo cluster IGM. 
   The galaxy colors, luminosity profile and SED are consistent with a sudden gas depletion in the last few hundred Myr.
   The SED fitting of the blobs constrain their ages in $< 400$ Myr. 
   }
   {}


%
\authorrunning{Fumagalli et al.}
\titlerunning{Constraining the ages of the fireballs in the wake of the dIrr galaxy VCC1217 / IC3418} 
\maketitle

\section{Introduction}

Recent studies of the Virgo Cluster (Boselli et al. 2008), Coma Supercluster (Gavazzi et al. 2010), Perseus Cluster (Penny \& Conselice, 2010)
and Shapley Supercluster (Haines et al. 2006) invoke ram pressure stripping (Gunn \& Gott 1972) as the responsible process for a
significant migration of galaxies from the Blue Cloud to the Red Sequence,  via suppression of the star formation due to gas ablation of low mass galaxies
in regions of high galactic density. The necessary ingredient of these "near-field Cosmology" approaches is that significant infall of 
low mass star forming objects exists along the filamentary structures onto the densest clusters. These galaxies have their star
formation truncated in a short timescale due to the interaction with the IGM.\\
Observations of stripped gas are frequent in the local Universe. Long narrow H$\alpha$ tails, stretching up to 150 Kpc, 
are reported in the Virgo Cluster (Kenney \&
Koopmann 1999), Abell 1367 (Gavazzi et al. 2001, Cortese et al. 2006) and Coma Cluster (Yagi et al. 2010) 
associated to infalling galaxies. There is however little evidence that star formation ignites in the stripped wakes, 
except in a few cases.
Cortese et al. (2007) discovers for the first time a complex
of star forming blobs in the trails of two spiral galaxies belonging to two clusters at z=0.2, and 
Yoshida et al. (2008) finds a unusual complex of blue "fireballs" associated to the Coma galaxy RB 199. 
These cases of star formation
in the wakes of stripped galaxies are remarkably similar to the hydrodynamical simulations by Kapferer et al. 2009.\\
Recently Hester et al. (2010) reports the discovery of a similar object in the Virgo cluster, associated 
to the dIrr galaxy VCC 1217 / IC 3418.
We have been independently studying the same system with deep LBT photometry in addition to public 
H$\alpha$ and GALEX-UV data, aimed at constraining the ages
of the galaxy and the fireballs via SED fitting.\\
Through the paper we assume a standard cosmology and a distance module of 31 mag for the Virgo Cluster A 
corresponding to a distance of 17 Mpc, as in Gavazzi et al. (1999).

\section{The data}
\label{The data}

VCC1217 has been observed by GALEX in March 2004 in the Near UltraViolet (NUV, 1750-2750 $\AA$) and in the Far UltraViolet 
(FUV, 1350-1750 $\AA$) bands, with an exposure time of $\approx 4000$ and $\approx 1600$ s respectively\footnote{Significantly shorter than quoted by Hester et al. (2010) who used
additional GALEX observations that are not yet public.}.  
A narrow H$\alpha$ band image has been taken at the ESO 3.6m telescope in 2004 (see Gavazzi et al. 2006).
Sources with a H$\alpha$ surface 
brightness higher than $\sigma_{min} = 3.16 \cdot 10^{-17} \rm erg/s/cm^2 / arcsec^2$ 
have been detected (2$\sigma$ of the background).
The galaxy is undetected at 21 cm, as reported by Hoffman et al. (1989), who used the Arecibo telescope 
to put a stringent upper limit of
$M_{\odot} \approx 3.46 \cdot 10^{6}$ on the HI gas in the
\begin{figure*}[ht!]
\centering
\includegraphics[width=16cm]{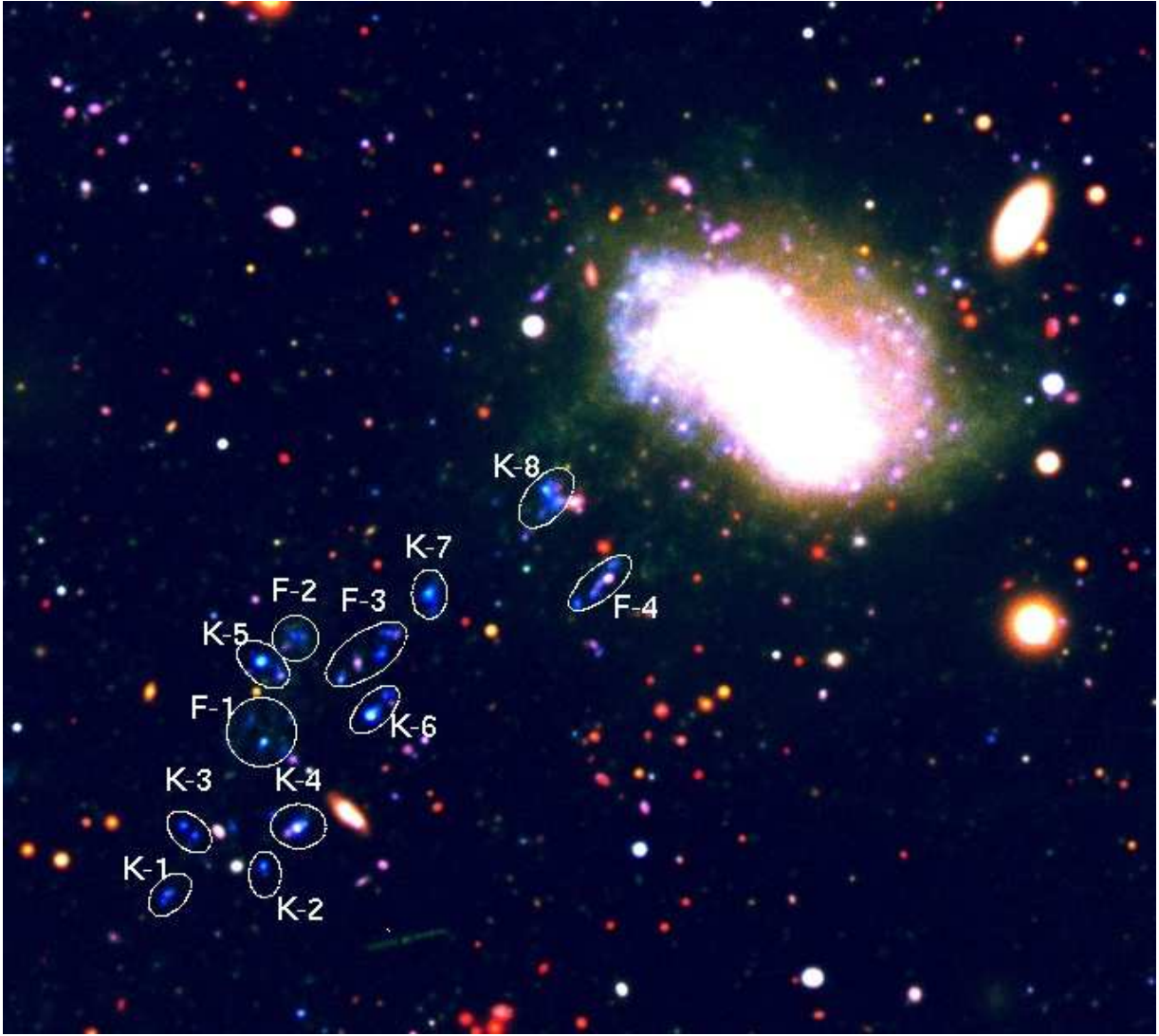}
\caption{High contrast RGB picture of VCC 1217 obtained from the LBT images (Uspec, g-SDSS, i-SDSS), 
highlighting the morphology and the color of the blobs.}
\label{Fig1}
\end{figure*}
\begin{table*}[hb!]
\caption{Observation Log}
\centering
{\footnotesize \begin{tabular}{lcllc}
\hline
\hline
Instrument & Filter & Seeing & Date (yy/mm/dd) & Exposure Time\\
\hline
\\
LBT & U-spec & 1.80 arcsec & 2008-02-01,02 & 25 x 240 sec \\
    &         &             & 2008-04-03,04  &               \\
LBT & g-SDSS & 1.35 arcsec &  2008-04-03,04 & 28 x 150 sec \\
    &         &             &  2009-02-22    &              \\
    &         &             &  2009-05-28    &              \\

LBT & r-SDSS & 1.47 arcsec & 2008-02-02    & 32 x 150 sec  \\
    &         &             & 2008-04-03,04    &               \\
    &         &             & 2009-05-28       &               \\

LBT & i-SDSS & 1.29 arcsec & 2009-02-01,02    & 34 x 240 sec  \\
    &         &             & 2008-04-03,04    &               \\

\\
ESO 3.6 & r Gunn & 1.13 arcsec &  2005-04-21 & 240 sec \\
ESO 3.6 & 692 & 1.13 arcsec & 2005-04-21 & 1800 sec\\ 
\\
GALEX & NUV & 4.95 arcsec & 2004-03-11 & 1597 + 1161 +1690 sec\\
GALEX & FUV & 4.05 arcsec & 2004-03-11 & 1597 sec \\
\\
\hline
\\
ESO 3.6 & EFOSC spectrometer &    & 2002-03-17 & 2400 sec \\
 \\
\hline
\end{tabular}
}
\label{Table1}
\end{table*}
\clearpage
\begin{figure*}[h!]	
\centering															     	  
a. \includegraphics[width=8cm]{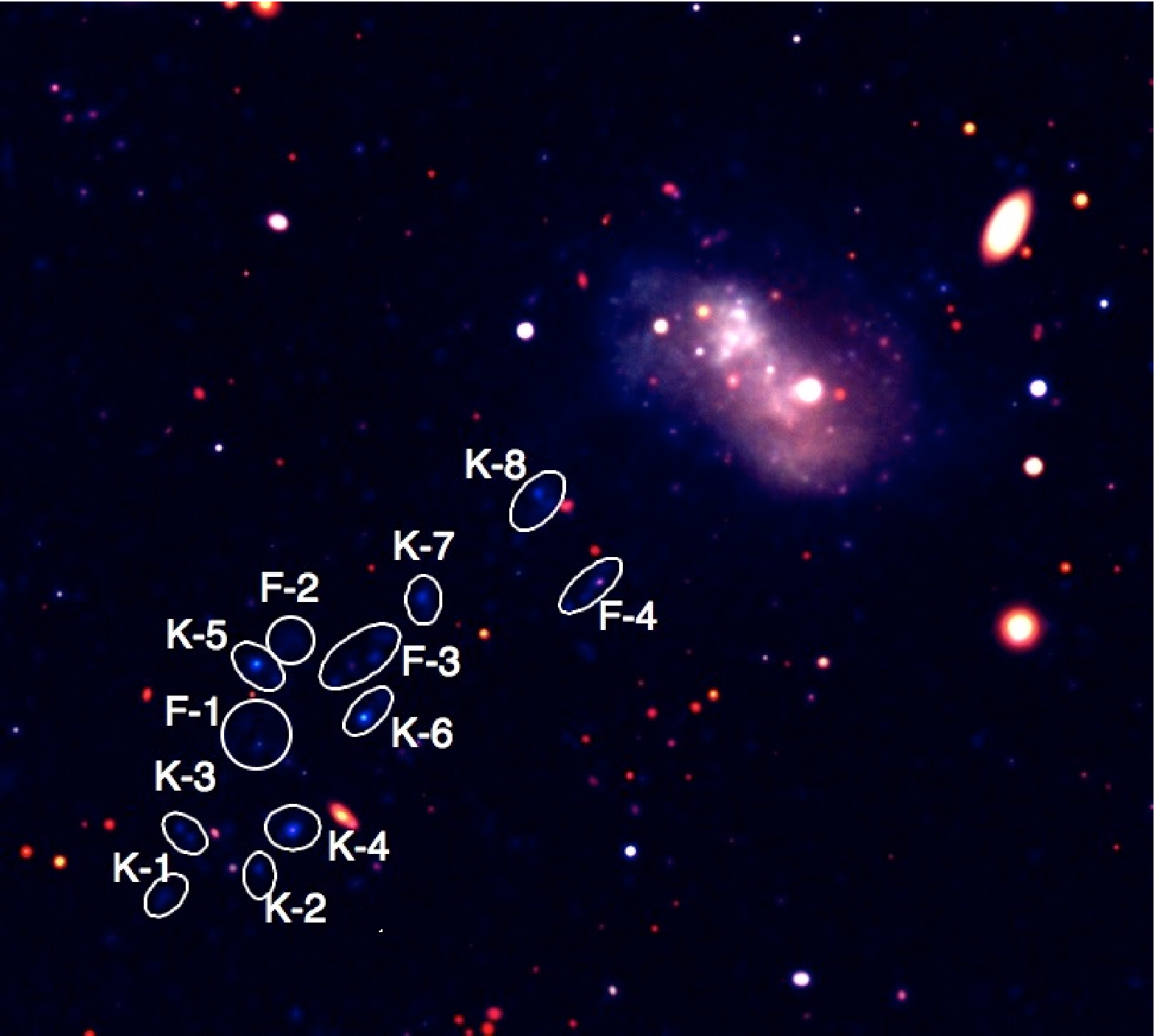}											     	  
c. \includegraphics[width=8cm]{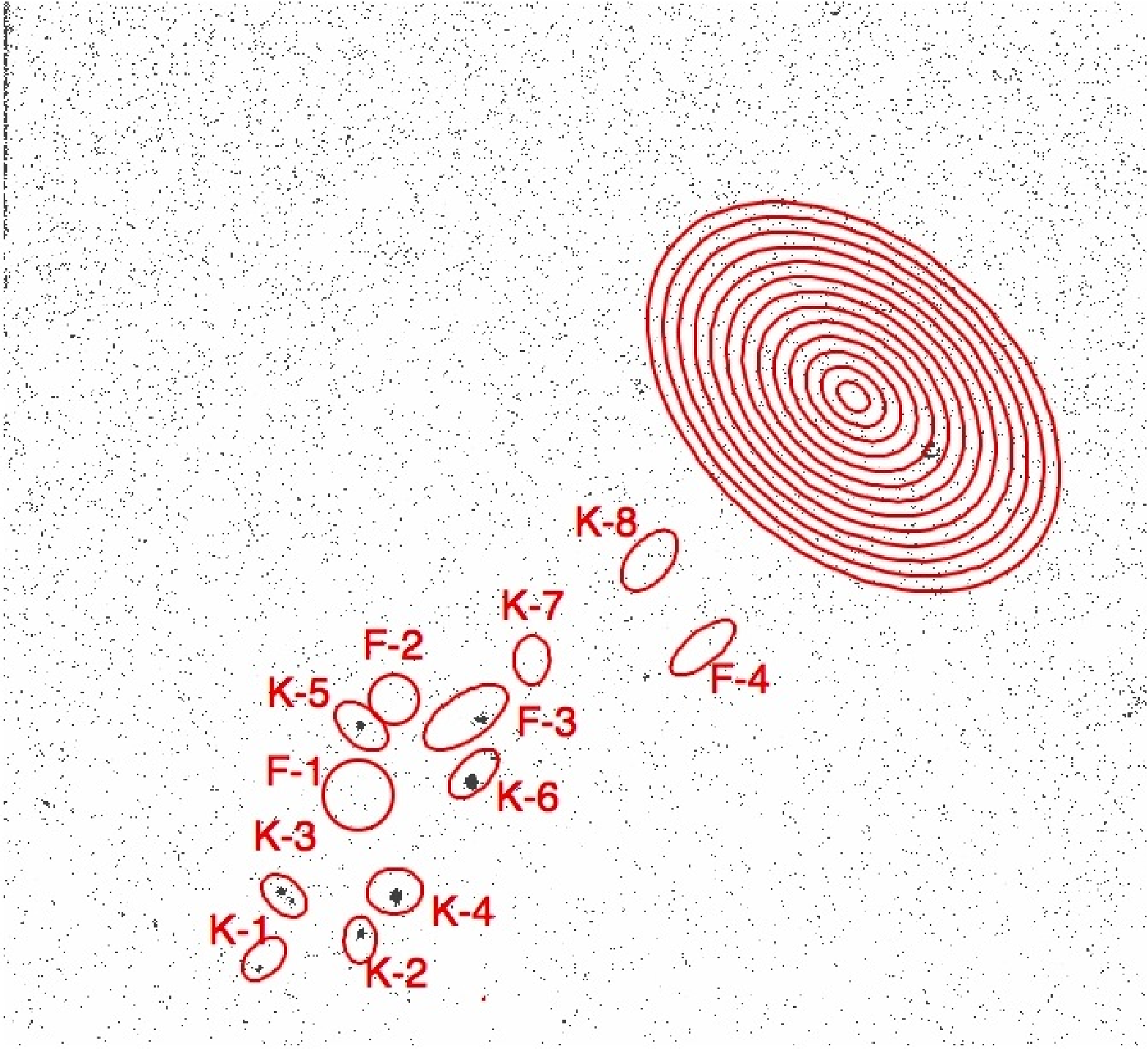}\\										     	  
b. \includegraphics[width=8cm]{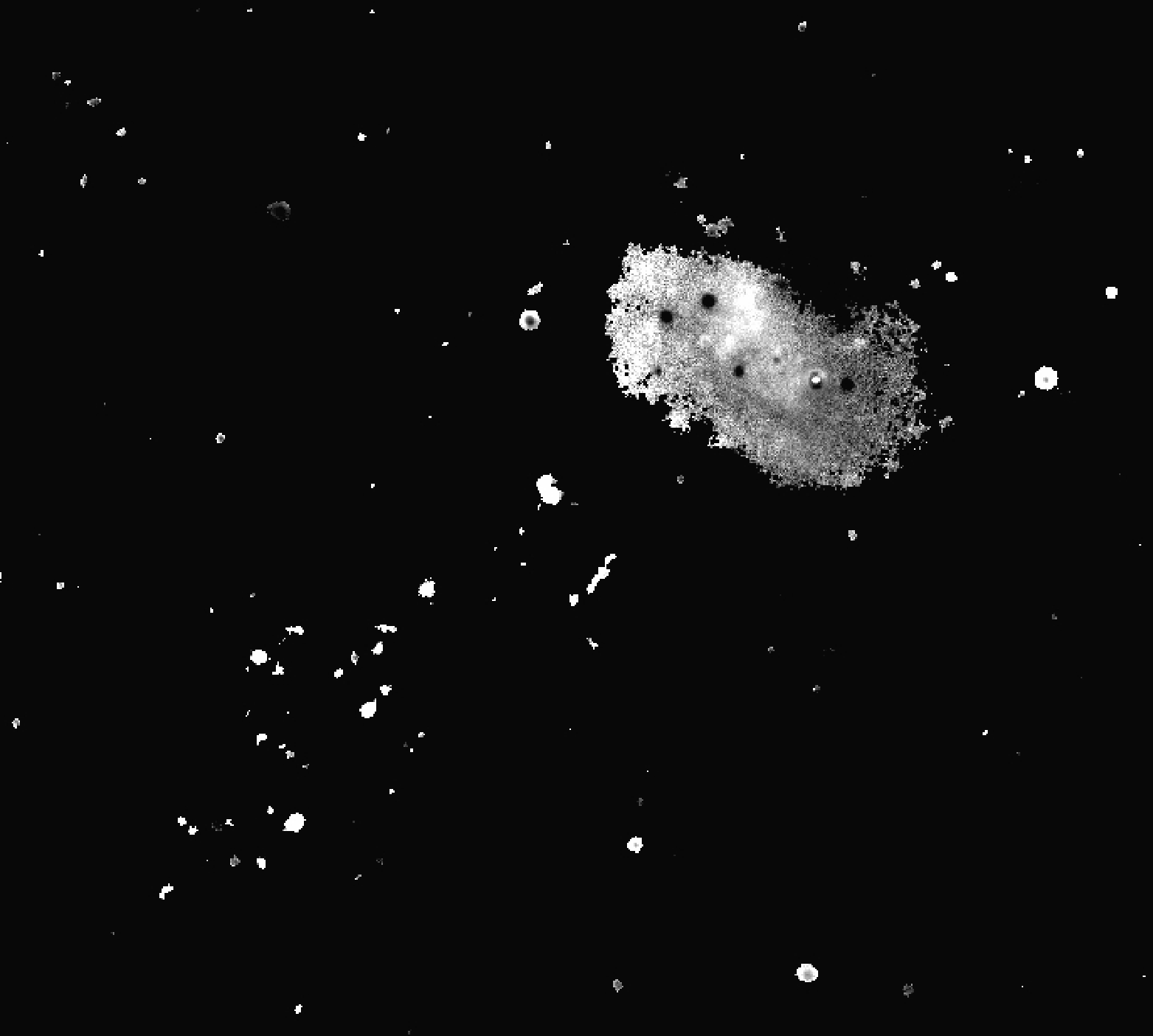}												     	  
d. \includegraphics[width=8cm]{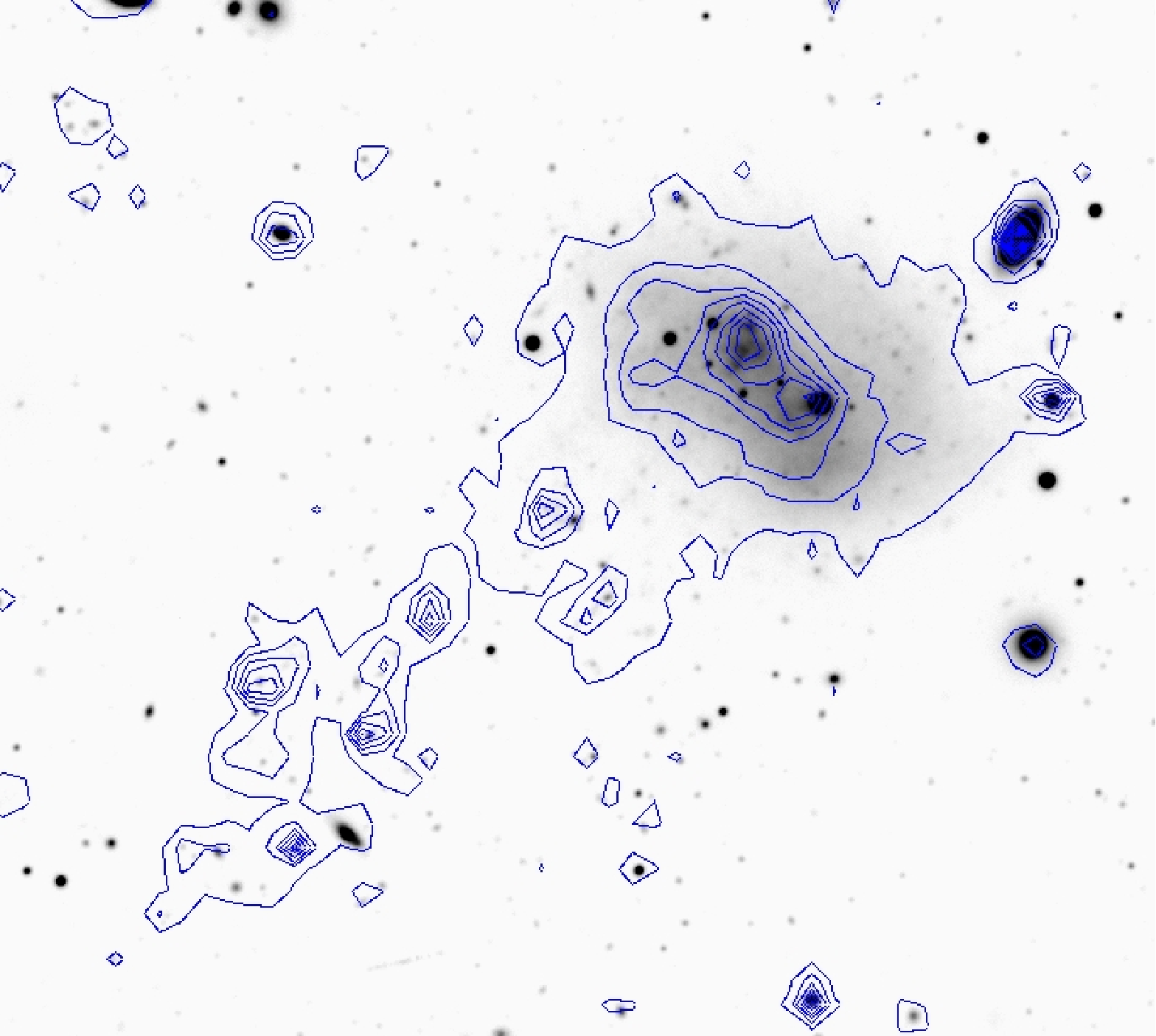}\\											     	  
\caption{ (a) Low contrast RGB picture highlighting the galaxy  structure and colors. (b) u-i image of the galaxy 
obtained from the ratio of the u and i images, each thresholded above $2\sigma$ of the sky (outer black regions).
In regions above the threshold the grey scale goes from white (u-i=1.5) to black (u-i=1.9) with increasing u-i index. (c) Superposed to the 
H$\alpha$ image, the regions on which the photometry
of the individual blobs has been evaluated and the concentric elliptical rings used to obtain the color 
profile of the galaxy (see Figure \ref{Fig4}) (d) NUV contours superposed
to a low contrast g image}	  
\label{Fig2}																     	  
\end{figure*}
\clearpage
\begin{table*}[h!]
\caption{Photometry of VCC1217 and its associated blobs.}
\centering
{\footnotesize \begin{tabular}{ccccccccc}
\hline
\hline
 ID & Distance &   u    &    g  &   r   &   i   &   NUV  &   FUV  \\
  & (arcmin) &   (mag)    &    (mag)  &   (mag)   &   (mag)   &   (mag)  &   (mag)  \\
{\tiny(1)}  & {\tiny(2)} & {\tiny(3)} & {\tiny(4)}  & {\tiny(5)}  &  {\tiny(6)} &  {\tiny(7)} &  {\tiny(8)}  \\
 \hline
 \\
         GAL  	 &      0    &	   15.98    $\pm$   0.03     &	 14.74	 $\pm$   0.02	 &    14.39   $\pm$   0.02   &	  14.19	  $\pm$   0.02	 &     17.36 $\pm$ 0.022 & 18.14 $\pm$ 0.06 \\ 
 \\
          K8     &    1.13   &	   20.79    $\pm$   0.12     &	 20.42	 $\pm$	 0.03	 &    20.41   $\pm$   0.04   &    20.16   $\pm$   0.05   &     21.05  $\pm$ 0.11 &  20.92  $\pm$ 0.19\\
          F4     &    1.28   &	   21.36    $\pm$   0.16     &   21.20   $\pm$   0.04	 &    21.17   $\pm$   0.05   &    21.01   $\pm$   0.08   &     21.63  $\pm$ 0.15 &  22.03  $\pm$ 0.29\\
          K7     &    1.83   &	   21.35    $\pm$   0.16     &   21.31   $\pm$   0.04	 &    21.71   $\pm$   0.06   &    21.98   $\pm$   0.11   &     21.39  $\pm$ 0.13 &  21.27  $\pm$ 0.21 \\
          F3     &    2.14   &	   20.92    $\pm$   0.13     &   20.76   $\pm$   0.04	 &    20.49   $\pm$   0.04   &    20.43   $\pm$   0.06   &     21.52  $\pm$ 0.14 &  21.86  $\pm$ 0.27\\
          K6     &    2.39   &	   20.83    $\pm$   0.13     &   20.94   $\pm$   0.04    &    20.74   $\pm$   0.04   &    20.96   $\pm$   0.07   &     21.23  $\pm$ 0.12 &  21.28  $\pm$ 0.22\\
          F2     &    2.40   &	   21.62    $\pm$   0.18     &   21.04   $\pm$   0.04	 &    21.07   $\pm$   0.05   &    20.91   $\pm$   0.07   &     21.75  $\pm$ 0.15 &  21.69  $\pm$ 0.26\\
          K5     &    2.55   &	   20.84    $\pm$   0.13     &   20.78   $\pm$   0.04	 &    20.82   $\pm$   0.04   &    20.90   $\pm$   0.07   &     20.90  $\pm$ 0.11 &  20.80  $\pm$ 0.18 \\
          F1     &    2.74   &	   21.00    $\pm$   0.14     &   20.58   $\pm$   0.03	 &    20.67   $\pm$   0.04   &    20.65   $\pm$   0.07   &     21.25  $\pm$ 0.12 &  21.41  $\pm$ 0.23\\
          K4     &    3.00   &	   20.94    $\pm$   0.13     &   21.04   $\pm$   0.04	 &    20.64   $\pm$   0.04   &    20.65   $\pm$   0.06   &     21.31  $\pm$ 0.13 &  21.17  $\pm$ 0.21\\ 
          K2     &    3.16   & 	   22.51    $\pm$   0.26     &	 22.83   $\pm$   0.08	 &    22.67   $\pm$   0.09   &    23.22   $\pm$   0.20   &     23.13  $\pm$ 0.27 &  22.92  $\pm$ 0.42  \\
          K3     &    3.30   & 	   21.93    $\pm$   0.20     &	 22.21   $\pm$   0.06	 &    22.03   $\pm$   0.07   &    22.43   $\pm$   0.14   &     22.19  $\pm$ 0.19 &  21.84  $\pm$ 0.27  \\
          K1     &    3.55   &     22.41    $\pm$   0.25     &	22.44	 $\pm$   0.07	 &    22.45   $\pm$   0.08   &    22.6    $\pm$   0.15   &     23.01  $\pm$ 0.26 &  22.37  $\pm$ 0.34 \\
 
 \\
\hline
\end{tabular}
}
\\
\small{(1) GAL = VCC1217, K-\# Knots, F-\# Filaments (2) Projected distance from the center of VCC1217 
(3) to (8) The photometric uncertainties are quadratic sum of the ZP error and Poisson error}
\label{Table2}
\end{table*}
\clearpage
galaxy.
Optical observations have been obtained at the LBT in different nights in the first semester of 2008, 
using the prime focus LBC cameras (http://lbc.mporzio.astro.it/),
with 4 filters: U$_{spec}$ and g,r,i in the SDSS system. 
The seeing ranged between 1.4 and 2 arcsec. Data has been reduced using the LBC standard pipeline by the LBC support team and final images are typically 
composed by stacks of $<$ 20 dithers.\\
The photometric calibration was performed cross-correlating the flux of 27 stars in the field with those in the SDSS,  
resulting in a zero points
with an error of less than 0.02 mags.
A summary of observations is found in Table \ref{Table1}. 
Magnitudes are given in the AB system throughout the paper.

\section{The Galaxy}
\subsection{Morphology and Photometry}
\label{Galaxy Morphology}
\begin{figure}[!t]
\includegraphics[width=9cm]{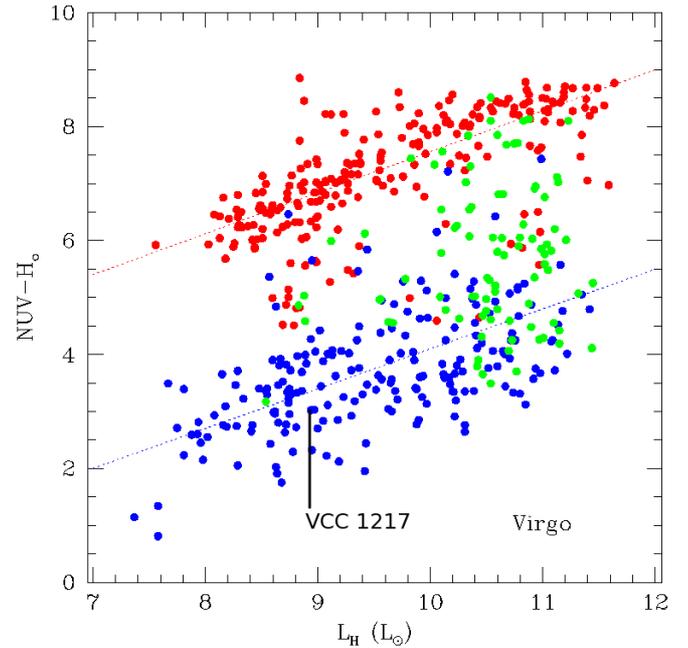}
\caption{NUV-H Color Luminosity diagram of the Virgo Cluster (from the GOLDMine database, Gavazzi et al. 2003). 
Objects are color coded according
to morphological classification: Red for Early Type Galaxies (Elliptical and S0), Blue for disk dominated 
Late Type Galaxies (Sc to Sd), Green for
Disk+Bulge Galaxies (Sa to Sb). Dashed lines represent the best fit to the Red and Blue sequences. 
The position of VCC1217 is highlighted.}
\label{Fig3} 
\end{figure}

VCC 1217 is a dIrr (Nilson et al. 1973) low surface brightness galaxy located near the center of the Virgo Cluster, 
approximately 1 degree South of M87 (0.31 Mpc 
projected distance) and with a redshift of 38 km/s, $\sim$ 1000 km/s lower than the mean redshift of Virgo Cluster A. 
At this location the emission from the hot IGM is at its peak intensity (Boehringer et al. 1994)
and the density of IGM is $\rho = 10^{-27} g / cm^{3}$ (Schindler et al. 1999).
The GOLDMine database (Gavazzi et al. 2003) reports for VCC 1217 an infrared luminosity 
of $H_{lum}=8.70 \cdot 10^{8} \rm L_{\odot}$ and a color $NUV-H = 3.02$ mag. 
With this characteristics VCC 1217 lies in the blue sequence of the Virgo Cluster (see Figure \ref{Fig3}).\\
VCC 1217 consists of a low surface brightness disk, whose structure is contaminated by several foreground stars.
The central brightest envelope, shaped as an elongated ellipse, contains
an excess of light in the North Eastern part of the object (Figure \ref{Fig2}a): this is the
brightest region of the galaxy and we will center all our subsequent radial analysis there.
A secondary peak is found more South West, offset by 15 arcsec. 

\begin{figure}[!t]
\begin{center}
\includegraphics[width=9cm]{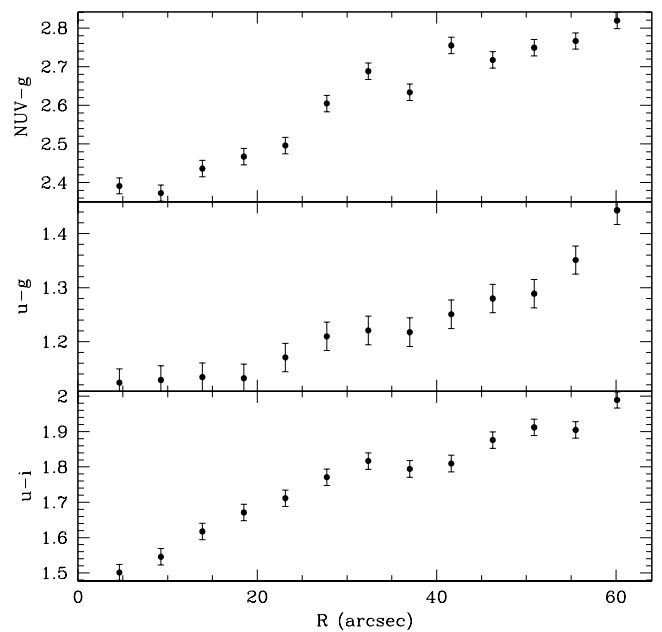}\\
\caption{ NUV-g (top), u-g (centre), u-i (bottom) color profile of VCC1217, obtained in the concentric annuli of Figure 2c.   
A significant positive color gradient is evident in all bands with increasing distance from the center up to 60 arcsec where contamination by the blue blobs is
null.}
\label{Fig4}
\end{center}
\end{figure}

\begin{figure}[htbp]
\includegraphics[width=9cm]{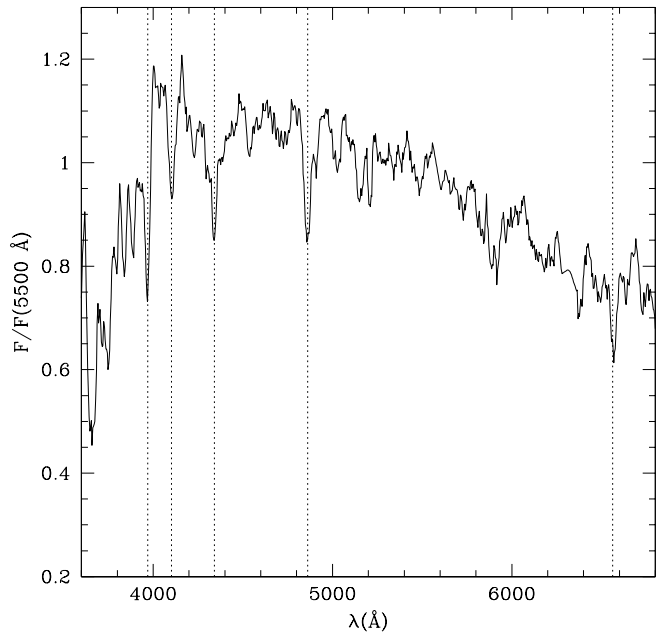} \\   
\caption{Spectrum of VCC1217 obtained in drift scan mode at ESO 3.6m telescope. The flux is normalized to 
F(5500 $\AA$). Dashed lines highlight the position of the Balmer series.}		         
\label{Fig5}			         
\end{figure}				         
The surface brightness profile has been evaluated  (after the masking the four most luminous foreground stars)
using a modified version of the ellipse task in IRAF, centering the ellipses on the 
brightest spot, and fitting the profile with an exponential law, with a scale length of 19.1 arcsec.\\
As revealed by both the color map u-i (Figure \ref{Fig2}b) and the RGB image (Figure \ref{Fig2}a), 
the brightest spot has a blue color of $u-i=1.5$ (see Figure \ref{Fig4}),
the secondary peak in the South West of the galaxy is redder ($U-i=1.87$).
We have performed a radial color analysis, 
integrating the u, g, i and NUV images on 12 elliptical annuli 
with major axis of 1 arcmin, an axis ratio of 1.5 and an inclination of the major axis of 50 degrees clockwise. 
Again we have masked the 4 more luminous stars superposed to the disk. 
The color profiles (Figure \ref{Fig4}) show a gradient of increasing color index 
by 0.4 mag with increasing distance from the center to 1 arcmin. 

\subsection{Spectroscopy}
A spectrum of the galaxy has been published by Gavazzi et al. (2004) and it is publicly available through GOLDMine
(Gavazzi et al. 2003).
It has been obtained at the ESO 3.6m telescope in drift scan mode, i.e. 
with the slit sliding over the whole galaxy area, thus representing
the mean spectral characteristics of the object.
We have smoothed the spectrum by 5 $\AA$ (Figure \ref{Fig5}) and measured the equivalent widths of the Balmer lines 
which result all 
stronger than 5 $\AA$ in absorption (as reported in Table \ref{Table3}), with no emission lines.
In particular, the H$\delta$ line has an EW=13.4$\AA$ (adopting the convention that positive EW mean absorption), i.e. 
stronger than the threshold in the diagnostic diagrams of k+a galaxies (e.g. Poggianti et al. 2004, Dressler et al 1999). 
k+a galaxies are interpreted to be Post Star-Burst (PSB) galaxies that underwent a sudden truncation of the star 
formation in the past 0.5-1.5 Gyr (Couch \& Sharples 1987).
\begin{table}[h!]
\caption{Spectroscopy of VCC1217}
\centering
{\footnotesize \begin{tabular}{cccc}
\hline
\hline
   	    & Wavelength      &    Continuum        &      EW     \\
   	    & $\AA$	      & F/F(5500$\AA$) &	    \\

\hline
\\
H$\epsilon$ & 3969.1	   & 1.02      &  8.0    \\
H$\delta$   & 4109.7	   & 1.18      &  13.4    \\
H$\gamma$   & 4339.8	   & 1.06      &  7.4    \\
H$\beta$    & 4861.0	   & 1.07      &  9.0     \\
H$\alpha$   & 6568.3	   & 0.78      &  6.2    \\
\hline
\end{tabular}
}
\label{Table3}
\end{table}
\section{The Fireballs}
A complex of faint blue knots and filaments extends from the galaxy in South East direction, up to 3.5 arcmin 
(17 Kpc). They are outstanding 
in both the GALEX data (Figure \ref{Fig2}d) and in the RGB image prepared with the u, g, i LBT images 
(Figure \ref{Fig1}, 
where the galaxy is saturated). We identify the brightest and clumpiest structures as "knots" (enumerating them from 
K-1 to K-8 East to West). Other structures with a lower surface 
brightness and visually more diffuse are labeled as "filaments", from F-1 to F-4. \\
Regions (highlighted in Figure \ref{Fig1} and \ref{Fig2}a, c) were selected
on the basis of their detection on the GALEX images (with a resolution $>$ 5 arcsec), 
even though the LBT images offer a better resolution.
This choice is dictated by the fact that we wanted to obtain for each feature the set of photometric measurements
over the full spectral range, from UV to i-band, necessary for a SED fitting analysis
(see Section \ref{SED Fitting}),
For instance, the region K-5 appears to be a bright knot in the NUV image (Figure 
\ref{Fig2} d), while it is resolved into two distinct blobs by the LBT. The same holds for filaments (see for instance F-3), 
which might consist of fainter knots connected by diffuse regions.
Although our regions do not exactly coincide with the ones in Hester et al. (2010)\footnote{The rationale for the small discrepancy between our regions and
the ones in Hester et al. (2010), beside nomenclature, is twofold: 1. our analysis initiated before the appearence of Hester et al. (2010); 
2. we positioned the regions on the basis of the NUV detection, but the fine tuning and the division into knots and filaments 
was aided by our high resolution LBT image}, 
the photometry in the FUV and NUV bands shows a general consistence.\\
\begin{figure}[b!]
\centering
\includegraphics[width=9cm]{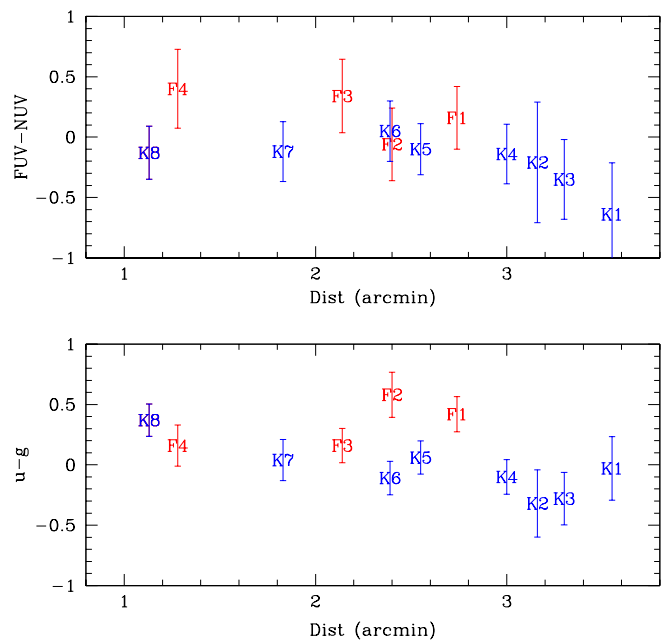}
\caption{Colors of the knots (blue) and filaments (red)} 
\label{Fig6}
\end{figure}
Figure \ref{Fig6} shows the fireballs in the two color differences FUV-NUV and u-g. 
The filaments appear marginally redder than the knots, having a mean color 
$\rm <FUV-NUV>_{F}= 0.21$ and $\rm <u-g>_{F}=0.31$, while knots have $\rm <FUV-NUV>_{K}= -0.21$ and $\rm <u-g>_{K}=-0.05$. 
Only one knot is as red as the filaments (K-6 in FUV-NUV, K-8 in u-g). 
Note that all the structures in the wake are much bluer than the galaxy (FUV-NUV=0.66 and u-g=1.4).
Moreover there is a slight dependence of the color of knots/filaments on the distance from the galaxy, i.e. the 
ones located farther away are $\approx 0.5$ mags bluer than the closest ones.\\
Various blobs display an H$\alpha$ emission, as shown in Figure \ref{Fig2}c and in Table \ref{Table4},
with luminosities ranging from $8 \cdot 10^{36} \rm erg\phantom{x}s^{-1} \rm$ (K-1) to $8 \cdot 10^{37} \rm erg\phantom{x}s^{-1} \rm$ (K-6), consistently
with the faint end of the HII luminosity function (Kennicutt et al. 1989).
The signal-to-noise ratio of the image (see Section \ref{The data}) is such that every star forming region 
with a flux larger than $2.51 \cdot 10^{-16} \rm erg\phantom{x}s^{-1} cm^{-2} str^{-1}$ can be detected 
(integrating 2$\sigma$  counts of the sky
on the typical dimension of a blob, 10 arcsec$^2$).
The H$\alpha$ emission is concentrated in the knots farther than 2.3 Kpc from the galaxy and only in one
filament (F-3), which was said to consist of smaller knots.
Consistently with the case of RB199 (Yoshida et al. 2008) the star formation resides in the 
most compact regions in the wake.
According to Kennicutt (1998), the ongoing SFR is evaluated to be of the order of $10^{-3/-4} M_{\odot} / yr$ 
in each blob (see Table \ref{Table4}), with a cumulative SF in the entire wake of $\approx 1.9 \cdot 10^{-3}  M_{\odot} / yr$. 
 
\section{SED Fitting}
\label{SED Fitting}

In order to reconstruct the star formation history of the blobs and of the galaxy, we 
generate spectral evolution models with the PEGASE2.0 code (Fioc \& Rocca-Volmerange 1997)
and perform a SED-fitting using GOSSIP (Franzetti et al. 2008).
For both the galaxy and the filaments/knots, which are analyzed separately, the procedure consists in:
\begin{itemize}
\item Giving as an input to PEGASE one or more Star Formation Histories, i.e. SFR(t)
\item Generating synthetic spectra at fixed times
\item Running in GOSSIP the SED-fitting between the synthetic spectra and the photometric points of the objects
\end{itemize}

\subsection{The Galaxy} 

We model the spectral evolution of the galaxy assuming a Salpeter IMF with an upper mass of $120 M_{\odot}$, 
a null initial metallicity and
a star formation history 'a la Sandage' (as reported in Gavazzi et al. 2003), with a discrete set of $\tau$ parameters.
Since we want to test the hypothesis of ram pressure stripping on the galaxy, 
in this phase we simulate the effect of the gas depletion including in the models a truncation of the star 
formation at a given characteristic time (Truncation Time, $t_{trunc}$).
The Star Formation History becomes:
$$
\rm 
SFR(t) = \left\{ \begin{array}{rl}
 \frac{t}{\tau^2} e^{-\frac{t^2}{\tau^2}} &\mbox{ if $t<t_{trunc}$} \\
  0 &\mbox{ if $t>t_{trunc}$}
       \end{array} \right.
$$
\begin{figure}[h!]
\begin{center}
\includegraphics[width=9cm]{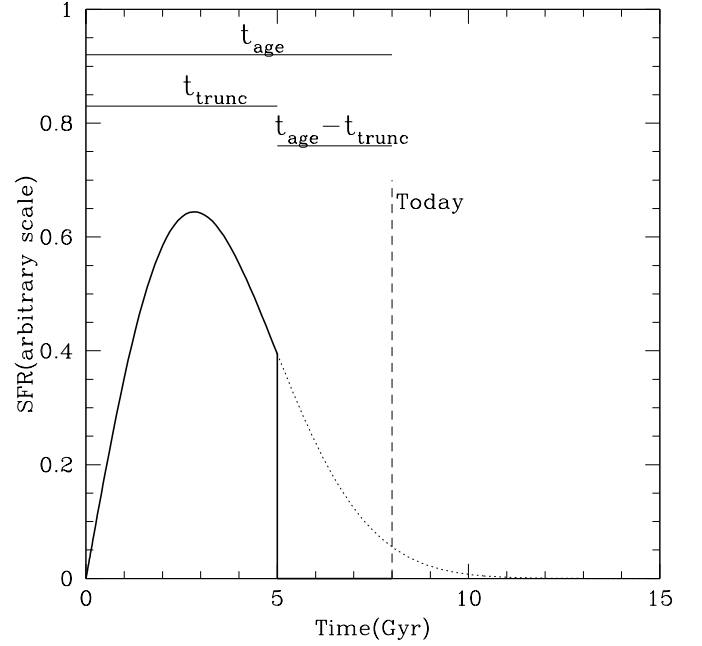}\\
\caption{A sample SFH from the library, with $\tau$=4 Gyr, $t_{age}$= 8 Gyr and $t_{trunc}$ = 5 Gyr}
\label{Fig7}
\end{center}
\end{figure}

\begin{figure}[t!]
\begin{center}
\includegraphics[width=4.5cm]{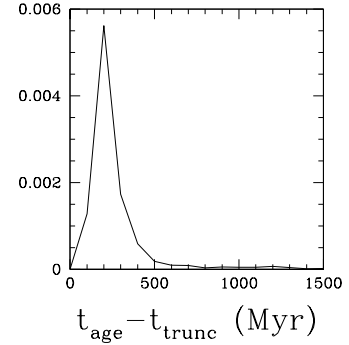}
\caption{Probability Distribution Functions of the $t_{age}-t_{trunc}$ parameter for VCC1217.
}
\label{Fig8}
\end{center}
\end{figure}

In Figure \ref{Fig7} we show a sample Star Formation History in the models library and highlight the different timescales referred to in the article: 
$t_{age}$ is the time from the onset of star formation to now, $t_{trunc}$ 
the time from the formation until the end of star formation activity, $t_{age}$-$t_{trunc}$ the period between the previous two.
The grid of parameters is built with $\tau$ ranging from 1 to 20 Gyr with 1 Gyr step and $t_{trunc}$ from 1 to 13 Gyr with 1 Gyr step, 
while $t_{age}$ spans from 0 to 13.5 Gyr with an step of 100 Myr, for a total of 35K spectra.
We remark that in the models building we don't include any fixed age for the start of star formation activity.
We don't include any correction for dust extinction, since for low mass galaxies it is negligible (see Figure 8 of Cortese et al. 2008 and 3-4 of Lee et al. 2009).
We run GOSSIP and evaluate the parameters and their probability distribution functions (PDFs).
We obtain that the $t_{age}$, the $t_{trunc}$ and $\tau$ are not well costrained.
However, we compute the probability distribution function of $t_{age}-t_{trunc}$, representing the lookback time at which
the truncation of star formation occurred, this parameter results very well constrained in $t_{age}-t_{trunc} = 200 ^{+90} _{-90} \rm Myr$. A halting of the 
star formation approximately 200 Myr ago is in accordance with the PSB signature in the galaxy spectrum (the PDF is given in Figure \ref{Fig8}).
The stellar mass of the galaxy evaluated from the normalization to the fit ($M_{star}=3 \cdot 10^{8} M_{\odot}$) 
turns out to be in fair agreement with the stellar mass evaluation from the optical data (Bell et al. 2007), 
$M_{star}=3.8 \cdot 10^{8} M_{\odot}$.

\begin{table}[t!]
\caption{
   Analysis:  (1) Distance from the galaxy in arcmin (2) H$\alpha$ Flux in units of 10$^{-16}$ erg/s/cm$^{2}$ (3) 
   Star Formation rate from the Kennicutt law, in units of 10$^{-3}$ M$_{\odot}$ /
   yr (4) Mass computed from the normalization to the SED fitting, in units of 10$^5$ M$_{\odot}$ 
   (5) Age (in Myr) for the best fit of a model with an exponential star formation rate. The error is computed from the probability distribution function.
   }
\centering
\begin{tabular}{c|c|cc|cc}   
\hline
\hline
 Name   & Distance & H$\alpha$ Flux & SFR    & Mass    &  Age       	     	     \\
       & (1)	  &  (2)	   & (3)    & (4)     & 	(5)  	     \\
\hline 					   				     	  							    
       &	  &		   &	    &	      & 		     \\
 K8    & 1.13	  &   - 	   &  -     &  3.74   &  130$^{+454}_{-150}$ 	\\
       &	  &		   &	    &	      & 		     \\
 F4    & 1.28	  &   - 	   &  -     &  2.45   &  620$^{+160}_{-90}$  	\\
       &	  &		   &	    &	      & 		     \\
 K7    & 1.83	  &   - 	   &  -     &  0.93   &   80$^{+23}_{-23}$   	\\
       &	  &		   &	    &	      & 		      \\
 F3    & 2.14	  & 7.76	   &  0.21  &  5.49   &  1400$^{+350}_{-150}$	\\
       &	  &		   &	    &	      & 		     \\
 K6    & 2.39	  & 0.25	   &  0.69  &  2.03   &  390$^{+87}_{-87}$   	\\
       &	  &		   &	    &	      & 		     \\
 F2    & 2.40	  &    -	   &  -     &  2.20   &  780$^{+192}_{-186}$ 	  \\
       &	  &		   &	    &	      & 		     \\
 K5    & 2.55	  & 6.92	   &  0.19  &  1.18   &  170$^{+39}_{-39}$   	  	\\
       &	  &		   &	    &	      & 		      \\
 F1    & 2.74	  &   - 	   &	    &  3.48   &  640$^{+248}_{-144}$ 	\\
       &	  &		   &	    &	      & 		      \\
 K4    & 3.00	  & 0.16	   &  0.10  &  3.79   &  740$^{+122}_{-122}$ 	\\
       &	  &		   &	    &	      & 		     \\
 K2    & 3.16	  & 6.22	   &  0.16  &  0.35   &  330$^{+123}_{-123}$ 	   \\
       &	  &		   &	    &	      & 		      \\
 K3    & 3.30	  & 8.31	   &  0.23  &  0.55   &  140$^{+52}_{-52}$   	   \\
       &	  &		   &	    &	      & 		     \\
 K1    & 3.55	  & 2.34	   &  0.06  &  0.40   &  260$^{+146}_{-146}$ 	   \\
        &	  &		   &	    &	      & 		     \\

\hline
\end{tabular}
\label{Table4}
\end{table}

%
%

\subsection{The Fireballs}
For the fireballs we compute with PEGASE2.0 a sample of synthetic spectra assuming a Salpeter IMF, a subsolar initial metallicity
and the following set of SFH models:
\begin{itemize}
\item Single burst $$\rm SFR(t) = \delta(0)$$
\item Constant Star Formation rate $$\rm SFR(t) = SFR_0$$
\item Exponential decrement of SFR rate
$$\rm SFR(t) = SFR_0 \cdot exp(-t / \tau)$$ with a finite set of $\tau$ parameters ($\tau =$ 10, 20, 40, 60, 80, 100,
200, 300, 400, 500, 600, 700, 800, 900, 1000, 1200, 1400, 1600, 1800, 2000 Myr).
\end{itemize}

\begin{figure}[t!]
\begin{center}
\includegraphics[width=9cm]{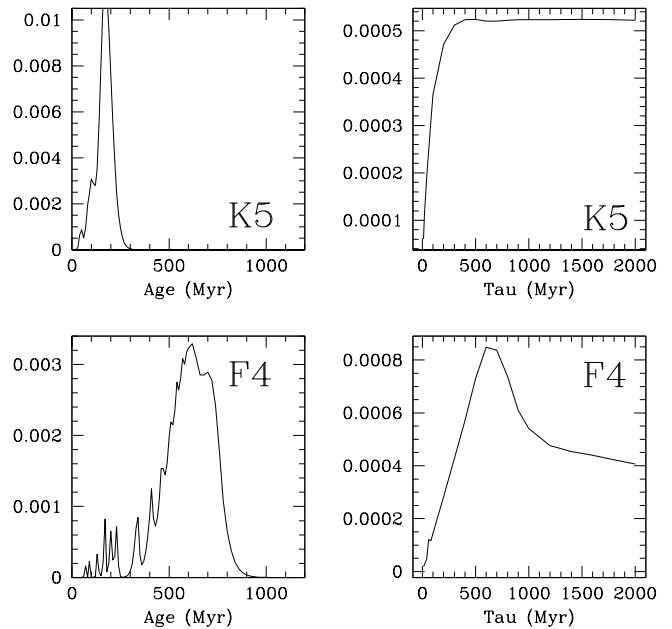}
\caption{Probability Distribution Functions of the Age and $\tau$ parameters for a sample knot and a sample blob.
}
\label{Fig9}
\end{center}
\end{figure}

\begin{figure}[t!]
\begin{center}
\includegraphics[width=9cm]{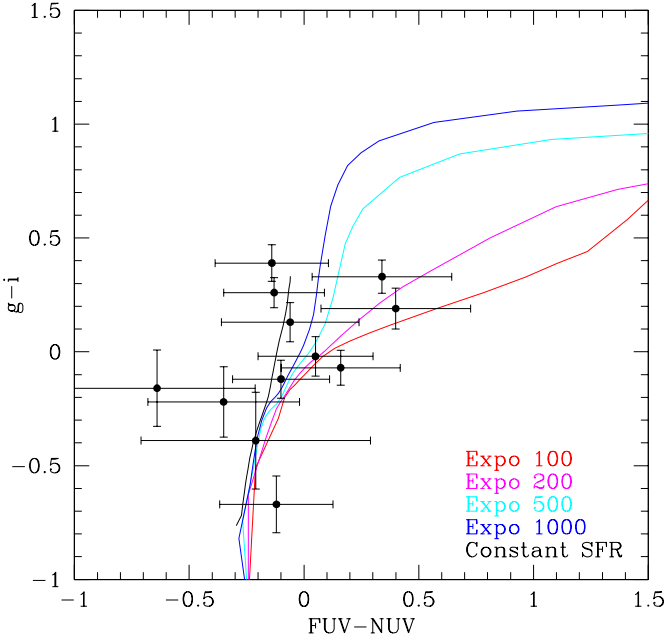}\\
\caption{Color-Color diagram of FUV-NUV and g-i for the blobs in the wake of VCC1217. The solid lines are the predictions of some SFH models: 
constant star formation rate and exponential star formation rate with $\tau$ 100, 200, 500 and 1000 Myr, color coded as in the legend.}
\label{Fig10}
\end{center}
\end{figure}

\begin{figure}[h!]
\begin{center}
\includegraphics[width=15cm]{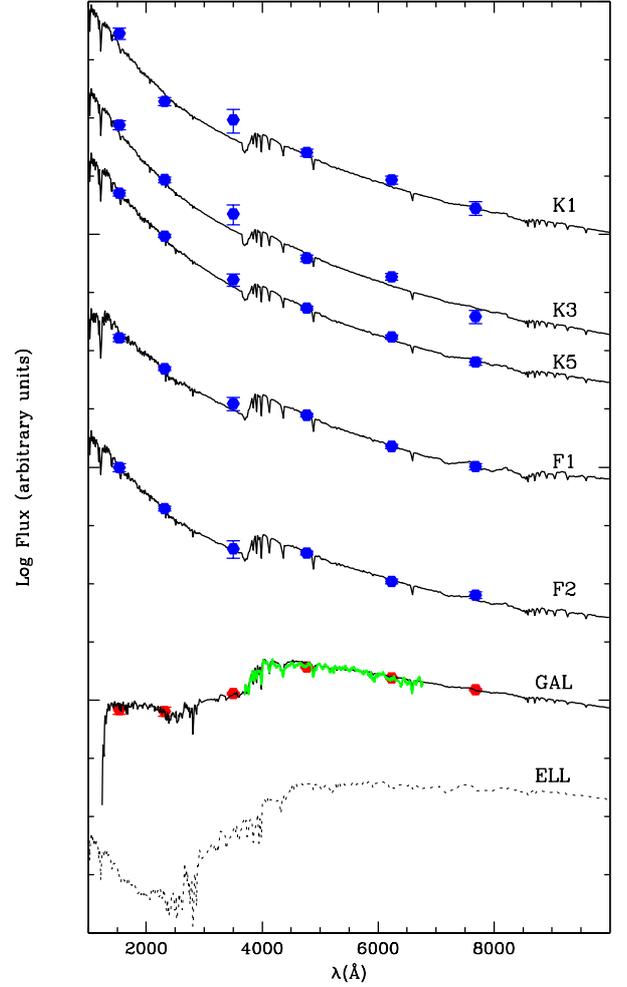}\\
\caption{Photometric points for the blobs (blue) and their best fit SEDs.
Photometric points (red), observed optical spectrum (green) for the galaxy VCC 1217 and best fit model with a 
truncated Sandage SFH. The dashed line represents for comparison the spectrum of an old quiescent elliptical galaxy
with a Sandage SFH with parameters Age = 13 Gyr and $\tau = 4$ Gyr}
\label{Fig11}
\end{center}
\end{figure}

Again we run the SED fitting with GOSSIP, and extract the best fits (Table \ref{Table4}) 
and evaluate their PDFs.
The stellar masses of the fireballs (derived from the normalization to the SED) range from $3.9\cdot10^{4}$ to 
$5.0\cdot 10^{5}$ M$_{\odot}$, which are typical dimensions of a Giant Molecular Cloud
 / HII region (Kennicutt et al. 1989). For such small objects we neglect the internal absorption.\\ 
We want to stress that we prefer not to improve the quality of the fit 
by replacement of our SFHs models (with at most one free parameter) with additional ad-hoc bursts.\\
Table \ref{Table4} contains the parameters of the SEDs. Most of the knots SEDs are well described by a simple exponential SFH with 
age $\rm t < 400$ Myr (excluding K4). We refer to Age as the time from the ingnition of the star formation to now.
Filaments appear to be all together older than blobs ($<Age>_{F} = 850$ Myr) with star formation activity stretched over a longer period of time. \\
While the age parameter is well constrained, as shown by the spiky shape of the PDFs in Figure \ref{Fig9} (left panels), not much can be
said about $\tau$, generally just a lower limit. From the PDFs of $\tau$ we can also conclude that a Single Burst Model 
(which corresponds to an exponential model with $\tau \rightarrow 0$) can be ruled out, while a model with a constant star formation rate
(which corresponds to an exponential model with $\tau \rightarrow \infty$) is consistent with the data, especially for the young knots.
The fact that for the knots $\tau$ is longer than the age indicates that star formation activity is still ongoing, in
accordance with the H$\alpha$ emission in most of the knots.\\
For such young objects it is impossible to constrain more precisely the star formation history by simply letting $\tau$ vary. 
This fact is illustrated in a color-color diagram (Figure \ref{Fig10}) where evolutionary tracks obtained with
various $\tau$ are plotted together with the data. At young ages different tracks are indistinguishable until the stellar population is more
evolved (compare with Fig. 13-14 of Yoshida et. al. 2008: the fireballs associated with RB199 are older than the ones in the present study).
Figure \ref{Fig11} collects some SEDs of the blobs and the galaxy, showing the quality of the fits. The features in the wake of VCC 1217 look all very similar, 
and extremely different from the galaxy.

\section{Ram pressure stripping}
Adopting the classical Gunn \& Gott (1972) criterion for ram pressure :
\begin{equation}
\rho_{IGM} v^2 \geq 2 \pi G \Sigma_{star} \Sigma_{gas}
\end{equation}
where $\rho_{IGM}$ is the intracluster density, $v$ the infall velocity and $\Sigma_{star,gas}$ the density of the 
star and gas components 
in the galaxy, and assuming an exponential profile for the stellar and gas component, the radius at which ram pressure 
becomes efficient
can be estimated as (Domainko et al, 2006):
\begin{equation}
R_{strip} = 0.5 R_{0} ln \left (\frac{GM_{star}M_{gas}}{v^2 \rho_{IGM} 2 \pi R_{0}^4} \right)
\end{equation}
and the stripped mass as: 
\begin{equation}
M_{strip} = M_{gas} \left(\frac{R_{strip}}{R_{0}}+1\right) exp\left(-\frac{R_{strip}}{R_{0}}\right)
\end{equation}
Adopting $M_{star}=3.8 \cdot 10^8 M_{\odot}$, $M_{star}/M_{gas} \approx 1$ (for late type galaxies, e.g. Boselli 2002),
$\rho = 6 \cdot  10^{-4} cm^{-3}$ for the IGM density of the Virgo cluster at the projected distance of 0.3Mpc from M87 
(Schindler et al. 1999), 
an infall velocity of 1000 km/s, and the typical scale length
computed in Section \ref{Galaxy Morphology}, 
Using these parameters we obtain that VCC 1217 is unable to retain its gas at any radius if subjected to ram pressure stripping ($R_{strip}=0.0 Kpc$) and therefore it results 
totally depleted of gas ($M_{strip} =  M_{gas}$).

\section{Discussion}
 
Although looking just at UV data\footnote{See \verb|http://www.galex.caltech.edu/media/images/|\\\verb|glx2010-02f_img01.jpg|} it can't be excluded 
that the system of blobs is the remnant of a dwarf irregular galaxy in some stage of merging with VCC1217, 
the LBT data resolves the tail feature into separated compact (at most filamentary) blobs, revealing that the morphology of the system is not consistent
with the merging scenario.
Several aspects of our analysis indicate that the ram pressure stripping picture is the most favorable one (in agreement with Hester and al. 2010), 
suggesting that VCC1217 has been recently 
stripped by the interaction with the Virgo Cluster IGM. The color profiles,
the spectroscopy and SED fitting of the galaxy all support the scenario consisting in a truncation of the star formation in the last few hundreds Myr.\\
The ensemble of blue knots and filaments stretching more than 17 Kpc South East of the galaxy is a remarkable feature. 
Their SEDs are consistent with very young stellar objects, born in the last few hundreds Myr, consistent with 
the timing of the gas depletion from the galaxy.\\ 
Similar objects have been observed in other clusters, but the phenomenon appears to be rare. 
Cortese et al (2007) report the discovery of two complexes of stellar tails 
and blue bright blobs associated with two spiral galaxies infalling in massive clusters at $z \approx 0.2$.
Yoshida et al (2008) analyze a complex of H${\alpha}$ emitting fireballs extending from the Coma cluster galaxy RB199, up to 80 Kpc.
Like VCC1217, also RB199 and 131124-012040 in Abell 1689 have k+a spectra, while 235144-260358 in Abell 2667 is still a star forming galaxy. 
Notice however the different scale between the objects in this study and the ones in the literature, both for the galaxies and the blobs.
RB199 is estimated to have a mass of 3-4 $\cdot 10^{9} M_{\odot}$, i.e. 10 times more massive than the one of VCC 1217. Yoshida et al.
compute a typical mass of the fireballs associated to RB199 of $\approx 10^{7-8} M_{\odot}$ and a total $L_{H\alpha}=2\cdot 10^{39} \rm erg\phantom{x}s^{-1} $, while 
the blobs in the present study range from 10$^{4.6}$ to 10$^{5.7}$ M$_{\odot}$ in mass and have a total H$\alpha$ luminosity of $2 \cdot 10^{38} \rm erg\phantom{x}s^{-1} $. 
Also the computed ages of the blobs are different, 500-1000 Myr for the complex 
stretching from RB199 and less than 400 Myr for most of the ones in the complex South West of VCC 1217.\\
The complex in the present study appears in conclusion to be a scaled down version of the one in Yoshida et al. (2008) because 
all its characteristic dimensions (galaxy mass, blobs masses, H$\alpha$ luminosities) are approximately 10 times smaller than the ones in RB199.\\
Besides the limited number statistics, this occurance might arise because the wake associated to VCC 1217 
would result too faint to be seen at the distance of Coma, but also because in a lower density environment,
such as Virgo compared to Coma, low mass galaxies are primarily affected by ram pressure (Bekki, 2009). \\
We note that this kind of events is extremely rare in the Universe, but the phenomenon happens at a variety of mass scales.
Different simulations (e.g. Tommesen et al 2010, Kapferer et al. 2009) have studied the impact of ram pressure in the distribution of gas in a galaxy,
producing mock observations in HI, H$\alpha$ (and X-rays) that are similar to the observed tails.
Various mechanisms are proposed for the H$\alpha$ emission in the wakes. Kenney et al. (2008) suggest that H$\alpha$ 
emission in the wake can be caused by thermal conduction from the IGM and turbulent shock heating.
In the current case however it is more likely to be associated with star formation, since the blobs are seen also
optically and have the SEDs typical of young stars complexes.\\
The simulations by Kapferer et al. (2009) conclude that turbulence in the wake can bring to gravitational instability and to the formation
of stars up to 100 Kpc behind the stripped galaxy. The observed pictures of VCC1217 (Fig.\ref{Fig2} a,b,c) are remarkably similar
to the mock observations (see Figures 9-12 in Kapferer et al. 2009). We can't compare directly the amount of new stars formed in the wake
since the simulations have been run assuming for the test galaxy a stellar mass of $2 \cdot 10^{10} M_{\odot}$ and a dark matter halo of 
$10^{12} M_{\odot}$, while VCC1217 is significantly smaller and the dependence on mass of the
combined effect of ram pressure, turbulence and gravitational instability has to be investigated further.\\
We quantify that after approximately 200 Myr from the stripping event the amount of new stars formed in the wake 
is $\approx 2.5 \cdot 10^{6} \rm M_{\odot}$ (i.e. the sum of the masses of the blobs, see Table \ref{Table4}), 1/100 of the whole mass of the stripped galaxy.
The non-detection in HI (Hoffman et al. 1989) is not surprising: assuming a residual gas mass similar to the mass of the blobs, it lies under the 
Arecibo detection limit. 

\section{Conclusion}
We propose that VCC 1217 has just interacted for the first time with the Virgo cluster, and underwent a sudden truncation of its gas content
due to ram pressure stripping. Turbulence in the wake can bring to gravitational instability and the densest parts of the wake into collapse,
with subsequent birth of stars. The analysis by Hester et al. (2010) is confirmed and reinforced by the spectroscopic inspection of the galaxy
and more notably by the deep LBT imaging. \\
It is already known that in the first stage of the interaction with the IGM a galaxy can form a tail of ionized gas stretching up to $> 50$ Kpc.
Examples of H$\alpha$ tails can be found in the Virgo Cluster (NGC 4522, Kenney \&
Koopmann 1999) and in Abell 1367 (97-079 and 97-073, Gavazzi et al. 2001; the BIG group, Cortese et al. 2006). 
A recent survey of the Coma cluster by Yagi et al. (2010) shows that almost all blue galaxies in the core of this cluster reveal tails and distorted
H$\alpha$ profiles when observed with a 10-m class telescope.\\
In spite of the high frequency of cometary H$\alpha$ structures associated with IGM-interacting galaxies, 
the inset of star formation in the wakes is less common and not fully understood. For instance within the Yagi et al. (2010) sample
only two objects show signs of star formation along the trail (GMP3016 and RB199).
VCC 1217 represents so far the closest and smallest known object with this unusual feature.

\section{Acknowledgments}
We thank Alessandro Boselli and Luca Cortese for the useful discussions and the LBT Survey Center (LBC) for carrying out the 
observation and for technical support during the reduction and analysis. This research has made use of the GOLDMine Database.
We acknowledge the anonymous referee for constructive criticism.

\clearpage

\end{document}